\documentclass{appolb}
\usepackage{graphicx}
\usepackage[utf8]{inputenc}

\begin{document}
\title{Hierarchical Cont-Bouchaud model}

\author{Robert Paluch and  Krzysztof Suchecki
\address{Center of Excellence for Complex Systems Research,\\Faculty of Physics, Warsaw University of Technology,\\Koszykowa 75, PL-00662 Warsaw, Poland}
\\ \vspace{10pt}
{Janusz A. Ho\l yst
\address{Center of Excellence for Complex Systems Research,\\Faculty of Physics, Warsaw University of Technology,\\Koszykowa 75, PL-00662 Warsaw, Poland}
\address{ITMO University, \\19, Kronverkskiy av., 197101 Saint Petersburg, Russia}
}}

\maketitle
\begin{abstract}
We extend the well-known Cont-Bouchaud model to include a hierarchical topology of agent's interactions.
The influence of hierarchy on system dynamics is investigated by two models.
The first one is based on a multi-level, nested Erd\H{o}s-R\'{e}nyi random graph and individual decisions by agents according to Potts dynamics.
This approach does not lead to a broad return distribution outside a parameter regime close to the original Cont-Bouchaud model.
In the second model we introduce a limited hierarchical Erd\H{o}s-R\'{e}nyi graph, where  merging of clusters at a level $h+1$  involves only  clusters that have merged at the previous level $h$ and we use the original Cont-Bouchaud agent dynamics on resulting clusters.
The second model leads to  a heavy-tail distribution of cluster sizes and relative price changes in a wide range of connection densities, not only close to the percolation threshold.
\end{abstract}

  
\section{Introduction}
The paradigm of investor's rationality and effectiveness was ousted in the second half of the 20th century by the behavioural economics.
A milestone was the work of Kahneman and Tversky written in 1979 \cite{kahneman}.
Since then, many studies were done and many models describing the stock market were developed \cite{scharfstein, grinblatt, trustinforeseeing, sieczka, sienkiewicz, szmaglinski, makowiec, wilinski}.
One of the first microscopic model was proposed by R. Cont and J.P. Bouchaud \cite{cont_bouchaud}.
It was suggested that the heavy tails observed in distributions of stock returns, corresponding to large fluctuations in prices, arise as a result of collective phenomena such as herding behaviour.
In the Cont-Bouchaud (CB) model, the groups (clusters) of investors make a collective decision during each time period and may choose either to buy the stock with the probability $P(\phi=+1) = a$, to sell it with the same probability $P(\phi=-1) = a$, or to stay inactive with the probability $P(\phi=0) = 1-2a$.
The clusters are formed according to the theory of Erd\H{o}s-R\'{e}nyi (E-R) random graphs: for any pair of nodes (agents), let $p=c/N$ be the probability that these nodes are linked together, where $N$ is total number of nodes and $c$ is a positive parameter, which represents the willingness of agents to form groups.
The demand/supply created by a cluster of investors is proportional to its size.
The price of an asset changes from one time step to another and the relative price change is proportional to the total excess demand.
Since for parameter $c$ close to and smaller than $1$, the cluster size distribution decreases asymptotically as a power law with exponential cutoff, the model can be solved analytically in the limit $2aN \approx 1$ according to the percolation theory.
Many modifications and extensions of the original CB model were proposed in past years: changing the topology from the random graph to a square ($d=2$), cubic ($d=3$) or hypercubic lattice ($d=4 \div 7$) \cite{crossover}, introducing moving agents \cite{monte_carlo_simulation} or fundamentalists \cite{fundamental}, introducing the dependence between the activity $a$ and the cluster's size \cite{selforganized} or between $a$ and a current price \cite{sharp_peaks} and much more \cite{spin_model, logperiodic_oscillations, multiscaling, asymmetries, kullmann, time_reversal, ising}.

All above studies assume only one level of the intimacy between traders (they are connected or they are not) and an infinitely strong interactions inside clusters.
These assumptions are hidden in oversimplified topologies like the random graph or the square lattice. Our aim was to adapt the main idea of the CB model to hierarchical topologies.
We expect that community of market investors, like other social systems, exhibit hierarchical structures.
An investor can divide others into different levels of "closeness" -- small group of friends or colleagues may influence their decisions in major way, including agreeing on common strategy, while the direct influence of the community as a whole may be small.
There is rarely collective behavior of a whole market, while this may happen much more often in close-knit groups.
Our model is a generalization of Cont-Bouchaud approach that considered only binary "deciding together" or "deciding separately" interactions between agents.
The nature of the interactions may be also interpreted as access to a common source of information, which translates into following the same trading strategy in an attempt to exploit that information.
Considering such interpretation one may argue that our models describe inefficient market, where dynamics can be sometimes dominated by large groups following single information sources, meaning it does not follow the aggregate of all information or may be biased.
Topological features of hierarchical systems have been observed in real networks \cite{ravasz}, and these kind of systems have been considered in several other studies, including social dynamics \cite{kondratiuk} and information propagation \cite{scirep}.
We developed two models, which use concept of hierarchy for extending the original CB model in two different ways.

\section{Hierarchical Erd\H{o}s-R\'{e}nyi graph}
\label{HER}
Consider a topology which consists of $N$ nodes which are grouped into many nested clusters.
The procedure of creating such a complex system is following:
\begin{enumerate}
	\item Link together each pair of nodes with a probability $p_1=c/N$. The result is $W_1$ clusters with the hierarchical degree $h=1$. The nodes belonging to the same cluster are nearest neighbours, so their mutual degree of neighbourhood is one.
	\item The existing clusters of hierarchical degree $h=1$ are treated as primary-level nodes which are linked together randomly with a probability $p_2=c/W_1$ per pair. The result is $W_2$ clusters with the hierarchical degree $h=2$, containing sets of clusters of degree $h=1$. If nodes $i$ and $j$ belong to the same cluster with the hierarchical degree $h=2$, but to different clusters of the hierarchical degree $h=1$, their mutual degree of neighbourhood is two.
	\item Repeat step 2 for following hierarchy levels $h$, until hierarchy level $h=H$ is reached, where all the nodes belong to only a few clusters. The number of iterations $H$ required is proportional to logarithm of system size $H \sim \ln N$. Assume all the clusters of level $H$ form a single cluster of hierarchical degree $H+1$ that contains all nodes.
\end{enumerate}
The resulting graph will be called a {\it hierarchical Erd\H{o}s-R\'{e}nyi graph} (HERG).  
\begin{figure}[htb]
	\centerline{
	\includegraphics[width=1\columnwidth]{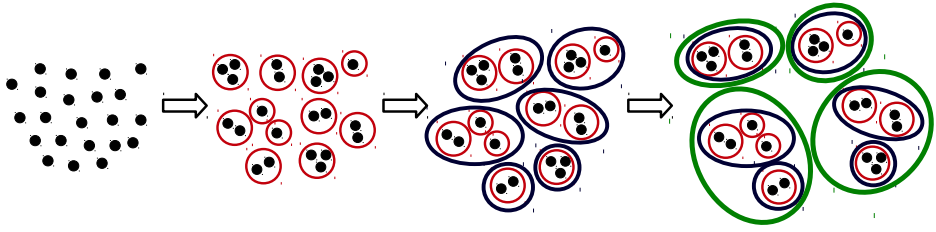}}
	\caption{The structure of hierarchical Erd\H{o}s-R\'{e}nyi graph (HERG). The nodes in red clusters are the nearest neighbours. Green clusters represent the neighbourhood of the third degree. Clusters are formed through connecting clusters of previous level at random, as in E-R random graph.}
	\label{Fig:scheme1}
\end{figure}

\section{Cont-Bouchaud model on hierarchical Erd\H{o}s-R\'{e}nyi graph with Potts interactions} \label{Potts}
We define a hierarchical Cont-Bouchaud model (HCB) in the following way.
Let nodes of HERG represent the stock traders.
At each time step, each agent has to take a decision about trading: buy ($\phi=1$), sell ($\phi=-1$) or do nothing ($\phi=0$), just as in the standard CB \cite{cont_bouchaud} model.

\begin{figure}[htb]
	\centerline{\includegraphics[width=0.45\columnwidth]{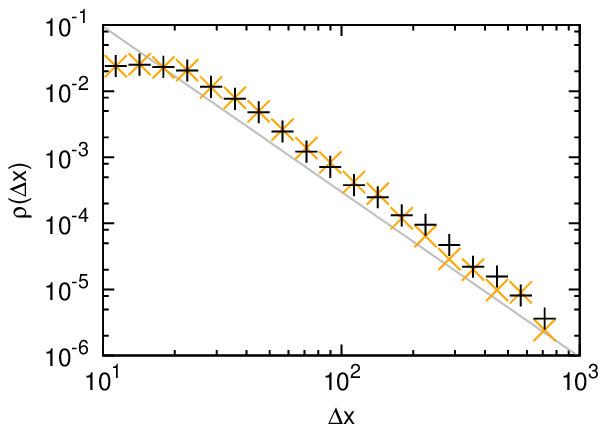}
	\includegraphics[width=0.45\columnwidth]{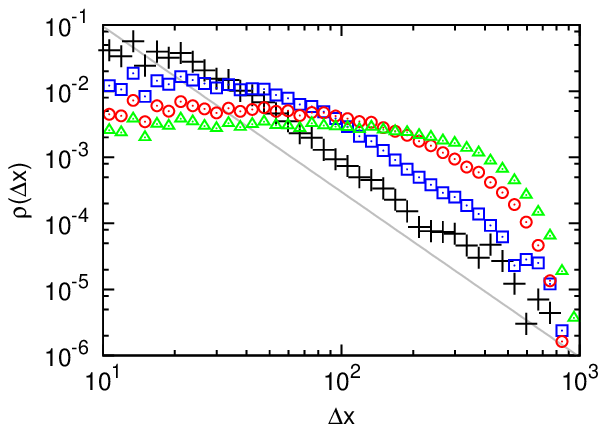}}
	\centerline{\includegraphics[width=0.45\columnwidth]{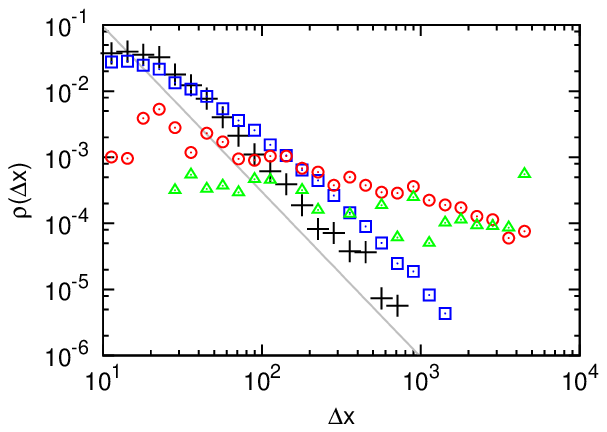}
	\includegraphics[width=0.45\columnwidth]{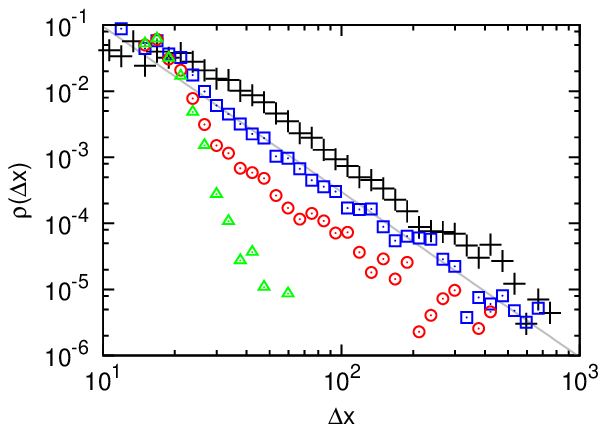}}
	\caption{Tails of return distribution densities for HCB model for $N=5000$, $c=0.99$, $\alpha=10^{-9}$, $\beta=1$, $J=1000$ and $B\approx 4.5$ (corresponding to $a=0.01$ in CB model).  Distributions in  HCB model for  this parameter  set   are the same as for the CB model  however deviations from that parameters result in fat tails disappearing. \emph{Top left:} comparison of HCB (black cross) with the original CB model (orange X) with the same parameters ($a=0.01$). \emph{Top right:} for decreasing field $B$: $\approx 4.5$ (black cross), $3.0$ (blue square), $1.5$ (red circle) and $0.0$ (green triangle). \emph{Bottom left:} for increasing $\alpha$: $10^{-9}$ (black cross), $0.1$ (blue square), $0.5$ (red circle) and $0.7$ (green triangle). \emph{Bottom right:} for decreasing $J$: $1000$ (black cross), $3.0$ (blue square), $2.0$ (red circle), $1.0$ (green triangle). The grey line is for the visual guidance only and it correspond to a power law with the  exponent $-2.5$.}
	\label{Fig:prices1}
\end{figure}

The agents are picked out at a random order and take decisions individually, but not independently, with probabilities corresponding to the probabilities of spin directions in Potts model in the canonical ensemble.
The probabilities are 
\begin{equation}
P(\phi_i) = \e^{-\beta E(\phi_i)}/Z, \label{prawdo}
\end{equation}
where $Z = \e^{-\beta E(-1)} + \e^{-\beta E(0)} + \e^{-\beta E(+1)}$ is the partition function and $\beta=1/k_BT$.
$E(\phi_i)$ is the spin energy corresponding to Potts Hamiltonian with interaction strengths $J_{ij}$ dependent on mutual degree of neighborhood $h(i,j)$ of interacting spins:
\begin{equation}
E(\phi_i) =  -\sum_{j \neq i} J_{h(i,j)} \delta(\phi_i,\phi_j) \:-\: B\delta(\phi_i,0), \label{energy}
\end{equation}
where $\delta(\phi_i, \phi_j)$ is the Kronecker delta, and the sum is over all agents $j \neq i$.
The parameter $B$ acts  as an external field which restrains (or enhances if it is negative) trading and therefore plays the same role as the parameter $a$ in CB model. In fact both parameters are related as $a=1/(e^{\beta B} + 2)$, so that when an agent is not influenced by others, then $P(+1)=P(-1)=a$.
The coupling factor $J_h = J\alpha^{h-1}$ depends on degree of neighbourhood $h$ and $J=J_1$ is the coupling constant which is equal to the coupling factor of the nearest neighbours.
The parameter $\alpha<1$ controls how quickly the interactions weaken with the degree of neighbourhood $h$.
At the beginning of each time step all agents are in a transient state \emph{null}, and do not influence other agent decisions.
During each time period the agents are appointed to make decisions in a random order, interacting according to Eq.\ref{prawdo} but only with agents in non-\emph{null} state that already made their decisions in this time step.
Then the return is calculated in a similar way as in CB model: 
\begin{equation}
\Delta x \propto \sum_{i=1}^N \phi_i(t).
\end{equation}
Let us note that there is no memory in agent states $\phi_i$ that are reset to the state \emph{null} after each time step.\\
We have investigated the model numerically for various parameters $\alpha$, $J$, $\beta$ and $B$, starting from the parameter set corresponding to the original CB model, where we assumed $\alpha=10^{-9}$, $J=1000$, $B\approx4.5$ (which correspond to $a=0.01$) with $\beta=1$.

When parameters are chosen so that the HCB model is replicating the original CB model, the return distribution is the same as resulting from the CB dynamics.
This behavior, along with results of changing parameters can be seen on Fig. \ref{Fig:prices1}, where the data have been aggregated over $r=100$ realization with $t=1000$ time steps each.
When the HCB model parameters deviate from those reflecting the CB model, the system stops working in  the ``percolation regime'' where the return distribution reflects the cluster size distribution and instead it works in the ``Potts regime'' where it follows the fluctuations arising from Potts interactions between agents.
We have not observed fat tails in the return distribution in the ``Potts regime'' for any analyzed parameter ranges (not only varying a single parameter from the CB parameter set).

\section{Limited hierarchical Erd\H{o}s-R\'{e}nyi graph}
In the Sect. \ref{HER} we introduced  a hierarchical Erd\H{o}s-R\'{e}nyi graph where all clusters of a hierarchical degree $h$ were considered as nodes at the next hierarchy level $h+1$ and they could be randomly linked to form clusters of that hierarchical degree.
Here we introduce a {\it limited hierarchical Erd\H{o}s-R\'{e}nyi graph} (LHERG).
The difference as compared to HERG model is that the merging of clusters at the level $h+1$ involves {\it only} clusters that have merged at the previous level $h$.
In effect the cluster growth is \emph{limited} only to clusters that succeed in finding partners at previous hierarchy level.\\

\begin{figure}[htb]
	\centerline{
	\includegraphics[width=1\columnwidth]{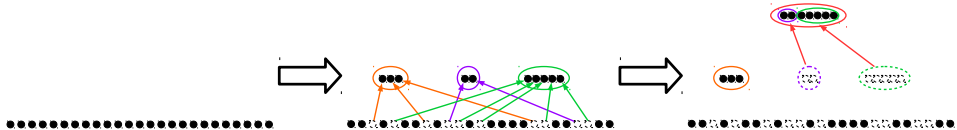}}
	\caption{Multilevel growth of clusters in LHERG model. Random E-R connections between single nodes (black) form clusters of level $h=1$ (orange, violet, green). Random connections between those clusters (single nodes do not participate, since they failed to connect) form clusters of level $h=2$ (red). Note that after the  second step, single nodes and the orange cluster are excluded from the further merging. We forget about the internal structure of clusters when we use the resulting disconnected clusters (orange, red and single nodes) for CB model in Sect. \ref{drugimodel}.}
	\label{Fig:scheme2}
\end{figure}
At the beginning, we consider the system of $N$ independent nodes that we treat as ``level zero'' clusters of size $1$.
The procedure is following:
\begin{enumerate}
	\item Link together each pair of nodes with the probability $p_1=c/N$. The result is $W_1$ clusters which advance to the next step. Not linked nodes are clusters of size one and they do not participate in further steps (and they are not included in $W_1$).
	\item Merge together each pair of clusters with the probability  $p_2=c/W_1$. The result of two clusters merging is a new cluster whose size is the sum of sizes of the merged clusters. New clusters ($W_2$) advance to the next step. The clusters which did not merge during this step do not advance and stop growing (and they are not included in $W_2$).
	\item Repeat the step 2 for clusters of successive levels until all clusters stop growing.
\end{enumerate}

\begin{figure}[htb]
	\centerline{
	\includegraphics[width=0.35\columnwidth]{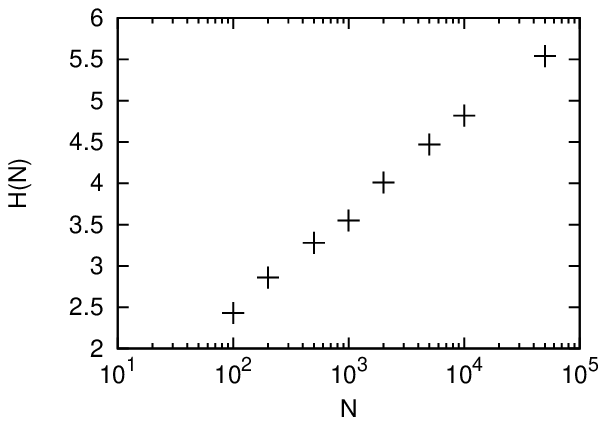}
	\includegraphics[width=0.35\columnwidth]{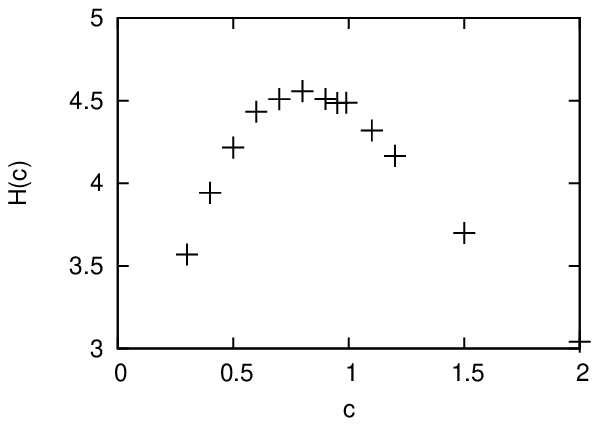}}
	\caption{The relations between the highest level of hierarchy $H$ and the size of the system $N$ (left picture) and between $H$ and the parameter $c$ (right picture) for LHERG model.}
	\label{Fig:hierarchies}
\end{figure}

Denote the average number of steps of the procedure of cluster growth by $H(c,N)$, which can be interpreted as the highest level of hierarchy of LHERG.
$H(c,N)$ is proportional to the logarithm on $N$ and has a maximum as a function of $c$ (this can be seen in Fig. \ref{Fig:hierarchies}).
Note that LHERG model results in many disconnected clusters (resulting from not connecting clusters that have been dropped out of the merging procedure), unlike in HERG model where all agents belong to a single top-level cluster.

\section{Cont-Bouchaud model on the limited hierarchical Erd\H{o}s-R\'{e}nyi graph} \label{drugimodel}
Now we  apply the dynamical rules of the standard CB model to the clusters obtained in the LHERG model.
This model is significantly different from the model presented in Sect. \ref{Potts}, and the differences can be expressed in 2 points:
\begin{enumerate}
\item The first model retains the hierarchical structure of the HERG clusters, while in the second it is forgotten after the clusters are determined through LHERG growth procedure. Note that LHERG results in multiple disconnected clusters.
\item The first model uses individual agent decisions and Potts interactions, while the second one uses collective random decisions by whole clusters as in the original CB model.  

\end{enumerate}

\begin{figure}[htb]
	\center
	\includegraphics[width=0.45\columnwidth]{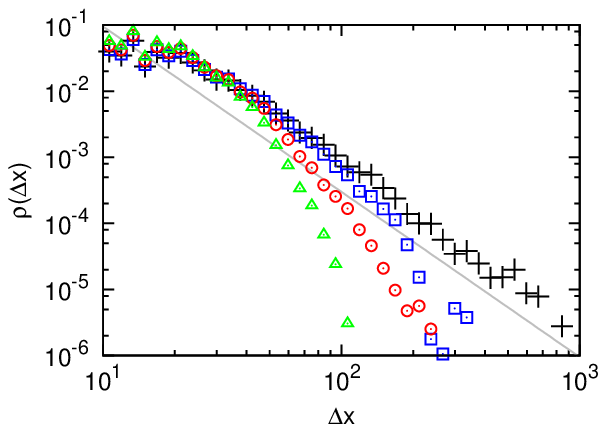}
	\includegraphics[width=0.45\columnwidth]{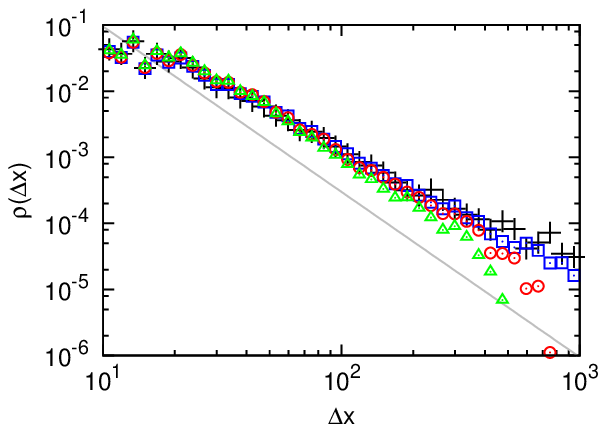}
	\caption{Comparison between the original CB model (left picture) and the CB model on the LHERG clusters (right picture). Distributions of price changes are computed for parameter $c=0.99$ (black cross), $0.9$ (blue square), $0.8$ (red circle), $c = 0.7$ (green triangle). The grey line is for visual guidance and has exponent $-2.5$. In the original CB model distribution quickly loses fat tail when $c$ goes away from percolation threshold, while in our LHERG model the power-law prevails.}
	\label{Fig:prices2}
\end{figure}

Numerical simulations show that our second model displays distributions of price changes (returns) with heavy tails for a wide range of parameter $c$, as seen in Fig. \ref{Fig:prices2}.
This means that assuming a hierarchical organization according to LHERG model, it is possible for CB model to explain fat tails of return distributions without making strict assumptions about the connection mechanisms between agents, such as operating at the percolation threshold.
We would like to note that while the distibution shape becomes insensitive to the parameter $c$, the order flow parameter $a$ changes return distributions just as in the standard CB model.\\

\begin{figure}[htb]
	\centerline{\includegraphics[width=0.45\columnwidth]{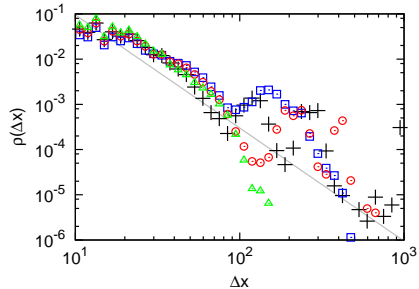}}
	\caption{Distribution of price changes for CB model on single realization of LHERG cluster. Distribution are computed for parameter $c=0.99$ (black cross), $0.95$ (blue square), $0.8$ (red circle), $0.5$ (green triangle).}
	\label{Fig:prices2r1}
\end{figure}

The CB model on LHERG clusters produces power-law fat tails when the results are averaged over multiple realizations.
When we keep to one, single realization, the distribution does not display power-law tail, and instead exhibits secondary peaks, that correspond to large clusters taking decisions to buy or sell (Fig. \ref{Fig:prices2r1}).
This means that the model can reproduce fat tails only when the topology is dynamical and changes in time, although not necessarily every time step.

\section{Conclusions}
We have introduced a model of hierarchical Erd\H{o}s-R\'{e}nyi graph, where nested clusters of successively higher hierarchies emerge from connecting lower-level components.
We have studied an extension of Cont-Bouchaud model, where we use interactions drawn from Potts model to decide whether agents buy or sell, allowing for different couplings between agent decisions, rather than an absolute correlation assumed in the CB original model.
We have used the hierarchical E-R graph as the topology, with couplings decreasing with a higher hierarchical degree of neighborhood.
The model does show a broad  return distribution only when working in the limit resembling the original CB model, meaning that the hierarchical structure of interactions does not result in power-law fluctuations when Potts interactions are used.
In the second approach the CB model has been  studied on clusters taken from the limited hierarchical Erd\H{o}s-R\'{e}nyi graph, where clusters can merge in following hierarchy levels only if they successfully make at least one connection in the current one.
If such a structure is kept dynamics, the cluster size distributions show on average power-law behavoir, therefore inducing  a power-law return distribution in CB model using this topology.
This happens for a wide range of average degrees $c$, not only when it is close to the E-R percolation threshold, therefore lifting one of the assumptions required for the model to explain fat tails in market fluctuation distributions, although necessitating dynamical cluster structure.

\section*{Acknowledgments}

The research leading to these results has received funding from the European Union Seventh Framework Programme
(FP7/2007-2013) under Grant Agreement No. 317534 (the Sophocles project) and from the Polish Ministry of Science and Higher Education Grant No. 2746/7.PR/2013/2. J.A.H. has been also partially supported by Russian Scientific Foundation, proposal \#14-21-00137  and by European Union COST TD1210 {\it KNOWeSCAPE} action.

%NOTE: While the DOI do not show in the bibliography style used (for making submission pdfs), they are present (as requested in Author Guide) in the bib file.

\bibliographystyle{polonica_mod}
\bibliography{Paluch_HCB_bibliography}

\begin{thebibliography}{10}

\bibitem{kahneman}
D.~Kahneman, A.~Tversky,
\newblock {\em Econometrica}
\newblock {\bf  47}, 263 (1979).

\bibitem{scharfstein}
D.S. Scharfstein, J.C. Stein,
\newblock {\em Am. Econ. Rev.}
\newblock {\bf  80}, 465 (1990).

\bibitem{grinblatt}
M.~Grinblatt, S.~Titman, R.~Wermers,
\newblock {\em Am. Econ. Rev.}
\newblock {\bf  85}, 1088 (1995).

\bibitem{trustinforeseeing}
J.A. Lipski, R.~Kutner,
\newblock {\em Acta Phys. Pol. A}
\newblock {\bf  123}, 584 (2013).

\bibitem{sieczka}
P.~Sieczka, J.A. Hołyst,
\newblock {\em Physica A}
\newblock {\bf  387}, 1218 (2008).

\bibitem{sienkiewicz}
A.~Sienkiewicz, T.~Gubiec, R.~Kutner, Z.R. Struzik,
\newblock {\em Acta Phys. Pol. A}
\newblock {\bf  123}, 615 (2013).

\bibitem{szmaglinski}
A.~Szmagliński,
\newblock {\em Acta Phys. Pol. A}
\newblock {\bf  123}, 621 (2013).

\bibitem{makowiec}
D.~Makowiec,
\newblock {\em Physica A}
\newblock {\bf  344}, 36 (2004).

\bibitem{wilinski}
M.~Wiliński, A.~Sienkiewicz, T.~Gubiec, R.~Kutner, Z.~R. Struzik,
\newblock {\em Physica A}
\newblock {\bf  392}, 5963 (2013).

\bibitem{cont_bouchaud}
R.~Cont, J.-P. Bouchaud,
\newblock {\em Macroeconomic Dynamics}
\newblock {\bf  4}, 170 (2000).

\bibitem{crossover}
D.~Stauffer, T.J.P. Penna,
\newblock {\em Physica A}
\newblock {\bf  256}, 284 (1998).

\bibitem{monte_carlo_simulation}
D.~Stauffer, P.M.C de~Oliveira, A.T. Bernardes,
\newblock {\em Int. J. Theor. Appl. Finance}
\newblock {\bf  2}, 83 (1998).

\bibitem{fundamental}
I.~Chang, D.~Stauffer,
\newblock {\em Physica A}
\newblock {\bf  264}, 294 (1999).

\bibitem{selforganized}
D.~Stauffer, D.~Sornette,
\newblock {\em Physica A}
\newblock {\bf  271}, 496 (1999).

\bibitem{sharp_peaks}
D.~Stauffer, N.~Jan,
\newblock {\em Physica A}
\newblock {\bf  277}, 215 (2000).

\bibitem{spin_model}
D.~Chowdhury, D.~Stauffer,
\newblock {\em Eur. Phys. J. B}
\newblock {\bf  8}, 477 (1999).

\bibitem{logperiodic_oscillations}
R.~B. Pandey, D.~Stauffer,
\newblock {\em Int. J. Theor. Appl. Finance}
\newblock {\bf  3}, 479 (2000).

\bibitem{multiscaling}
F.~Castiglione, D.~Stauffer,
\newblock {\em Physica A}
\newblock {\bf  300}, 531 (2001).

\bibitem{asymmetries}
I.~Chang, D.~Stauffer, R.~B. Pandey,
\newblock {\em Int. J. Theor. Appl. Finance}
\newblock {\bf  5}, 585 (2002).

\bibitem{kullmann}
L.~Kullmann, J.~Kertész,
\newblock {\em Int. J. Mod. Phys. C}
\newblock {\bf  12}, 1211 (2001).

\bibitem{time_reversal}
I.~Chang, D.~Stauffer,
\newblock {\em Physica A}
\newblock {\bf  299}, 547 (2001).

\bibitem{ising}
L.R. Da~Silva, D.~Stauffer,
\newblock {\em Physica A}
\newblock {\bf  294}, 235 (2001).

\bibitem{ravasz}
E.~Ravasz, A.-L. Barab\'asi,
\newblock {\em Phys. Rev. E}
\newblock {\bf  67}, 026112 (2003).

\bibitem{kondratiuk}
P.~Kondratiuk, J.A. Hołyst,
\newblock {\em Acta Phys. Pol. A}
\newblock {\bf  121}, B67 (2012).

\bibitem{scirep}
A.~Czaplicka, J.A. Hołyst, P.M.A. Sloot,
\newblock {\em Scientific Reports}
\newblock {\bf  3} (2013).

\end{thebibliography}
\end{document}